\documentclass[12pt,preprint]{aastex}

\begin{document}

\title{Triggering Collapse of the Presolar Dense Cloud Core 
and Injecting Short-Lived Radioisotopes with a Shock Wave. 
II. Varied Shock Wave and Cloud Core Parameters}

\author{Alan P.~Boss and Sandra A.~Keiser}
\affil{Department of Terrestrial Magnetism, Carnegie Institution,
5241 Broad Branch Road, NW, Washington, DC 20015-1305}
\email{boss@dtm.ciw.edu, keiser@dtm.ciw.edu}

\begin{abstract}

 A variety of stellar sources have been proposed for the origin of
the short-lived radioisotopes that existed at the time of the 
formation of the earliest Solar System solids, including Type II
supernovae, AGB and super-AGB stars, and Wolf-Rayet star winds.
Our previous adaptive mesh hydrodynamics models with the FLASH2.5 
code have shown which combinations of shock wave parameters are
able to simultaneously trigger the gravitational collapse of a
target dense cloud core and inject significant amounts of shock
wave gas and dust, showing that thin supernova shocks may be uniquely
suited for the task. However, recent meteoritical studies have
weakened the case for a direct supernova injection to the presolar
cloud, motivating us to re-examine a wider range of shock wave 
and cloud core parameters, including rotation, in order to better
estimate the injection efficiencies for a variety of stellar 
sources. We find that supernova shocks remain as the most promising
stellar source, though planetary nebulae resulting from AGB star
evolution cannot be conclusively ruled out. Wolf-Rayet star winds,
however, are likely to lead to cloud core shredding, rather
than to collapse. Injection efficiencies can be increased when the
cloud is rotating about an axis aligned with the direction of the shock 
wave, by as much as a factor of $\sim 10$. The amount of gas and dust 
accreted from the post-shock wind can exceed that injected from the shock
wave, with implications for the isotopic abundances expected for
a supernova source. 

\end{abstract}


\section{Introduction}

 The origin of the short-lived radioisotopes (SLRIs) that were present
during the formation of the earliest solids in the protoplanetary disk
that formed our planetary system remains as a largely unsolved mystery 
(Dauphas \& Chaussidon 2011; Boss 2012). SLRIs such as $^{60}$Fe and 
$^{26}$Al, with half-lives of 2.62 Myr and 0.717 Myr, respectively,
have been ascribed to nucleosynthesis in Type II (core collapse) 
supernovae, asymptotic giant branch (AGB) stars, or, for $^{26}$Al, in 
addition to these stellar nucleosynthetic sources, to irradiation 
of the inner edge of the solar nebula by the protosun. Having a
nearby supernova as a source of SLRIs requires that the Sun formed
in a massive cluster with at least 1000 stars, so that at least one
of the stars is likely to have a mass greater than 25 $M_\odot$ (e.g., 
Dukes \& Krumholz 2012). Most stars are born in such clusters, making
such a birth site a likely possibility for the Sun. Intermediate-mass
($\sim$ 5 $M_\odot$) AGB stars are possible sources for these SLRIs
(Huss et al. 2009). Super-AGB stars, with masses in the range
of 7 to 11 $M_\odot$, are also possible sites for synthesizing the 
SLRIs $^{60}$Fe and $^{26}$Al, as well as $^{41}$Ca, with a half-life
of 0.102 Myr (Lugaro et al. 2012). Due to their age, however, AGB stars
are unlikely to be found in regions of ongoing star formation: 
Kastner \& Myers (1994) estimated the chance that a mass-losing
AGB star would pass through a molecular cloud within 1 kpc of the
Sun in the next million years was only $\sim 1 \%$. However, they
also estimated that over a molecular cloud lifetime, the chances
of an AGB star interaction could be as high as $\sim 70\%$.

 Boss et al. (2010) studied the two-dimensional (2D) interactions
of shock waves with a variety of speeds with a $\sim 2 M_\odot$ cloud
core, finding that shock speeds in the range of $\sim 5$ km s$^{-1}$
to $\sim 70$ km s$^{-1}$ were able to simultaneously trigger cloud
collapse and inject significant amounts of shock wave gas and dust.
Boss \& Keiser (2010) showed that the injection efficiency  
depended sensitively on the assumed shock width and density,
and suggested that thin supernova shocks were therefore preferable
to the thick planetary nebula winds outflowing from AGB stars.
Gritschneider et al. (2012) modeled the 2D interaction of higher
speed (97.5 or 276 km s$^{-1}$) supernova shock waves with a 
10 $M_\odot$ cloud core, finding that simultaneous triggered 
collapse and SLRI enrichment was still possible in that case. Boss \& 
Keiser (2012) extended their 2010 models to three dimensions (3D),
revealing the 3D structure of the Rayleigh-Taylor (R-T) fingers that
form during the shock-cloud collision and which constitute the
injection mechanism.

 Other injection scenarios have been investigated as well.
Ouellette et al. (2007, 2010) studied the 2D injection of SLRIs
from a supernova into a nearby, previously-formed protoplanetary
disk, showing that injection of shock wave gas could not occur, but
that refractory grains with sizes of $\sim 1 \mu$m or larger could 
be injected. Pan et al. (2012) studied the 3D interaction of a
supernova remnant with the gaseous edge of the HII region created
by a supernova's progenitor star, finding that the supernova
ejecta could contaminate enough of the molecular cloud gas to
explain the observed levels of some Solar System SLRIs.

 Recent research has cast significant doubt on the initial amount of
$^{60}$Fe in the solar nebula. The initial ratio $^{60}$Fe/$^{56}$Fe 
was once thought to be high enough to require a nearby supernova.
However, new whole rock analyses on a range of meteorites have 
lowered this initial ratio to an apparently Solar-System-wide, 
uniform ratio of $1.15 \times 10^{-8}$ (Tang \& Dauphas 2012). Other 
studies have also found significantly lower ratios ($3 \times 10^{-9}$:
Moynier et al. 2011), or cast severe doubt on the much higher ratios 
previously found (Telus et al. 2012). Given that the interstellar medium 
(ISM) has a $^{60}$Fe/$^{56}$Fe ratio of $\sim 2.8 \times 10^{-7}$,
Tang \& Dauphas (2012) suggest that the meteoritical $^{60}$Fe may
have originated from the ISM rather than a nearby supernova, though
this would require the presolar cloud to have formed from a molecular
cloud that had not been injected with any fresh $^{60}$Fe for $\sim$
15 Myr, allowing the $^{60}$Fe to decay for $\sim$ 5 half-lives,
reducing the ISM $^{60}$Fe/$^{56}$Fe ratio by a factor of $\sim$ 30 to
the meteoritical value. Tang \& Dauphas (2012) then suggest that the
solar nebula's $^{26}$Al may have arisen from the $^{26}$Al-rich,
$^{60}$Fe-poor wind of a pre-supernova massive ($> 30 M_\odot$) star, 
i.e., a Wolf-Rayet (WR) star (e.g., Gaidos et al. 2009; Tatischeff 
et al. 2010). Tang \& Dauphas (2012) predict that if their suggestion 
is correct, calcium, aluminum-rich inclusions (CAIs), which have some 
variability in their initial $^{26}$Al/$^{27}$Al ratios (e.g., Krot 
et al. 2012), should have uniform initial $^{60}$Fe/$^{56}$Fe ratios.

 Gounelle and Meynet (2012) also proposed that the SLRIs came 
from a mixture of supernovae (for the $^{60}$Fe) and massive star winds 
(for the $^{26}$Al) from several generations of stars in a giant 
molecular cloud (GMC). Their scenario proposes that a first generation
of massive stars became supernovae and ejected $^{60}$Fe and $^{26}$Al 
into the GMC. After 10 Myr, a second generation of massive O stars formed, 
with the wind from one of them sweeping up GMC gas into an expanding shell
of gas rich with $^{26}$Al from its own outflow. Because of the shorter 
half-life of $^{26}$Al compared to $^{60}$Fe, after 10 Myr the 
$^{60}$Fe from the first generation of massive stars has decayed much 
less than the original $^{26}$Al, which has essentially disappeared. 
Gounelle and Meynet (2012) then propose that the Sun formed from
the expanding shell surrounding the massive star's HII region, before
the massive star itself becomes a supernova. It is unclear, however,
whether a single GMC, with a typical mass of $\sim 10^6 M_\odot$, could 
have two distinct episodes of massive star formation, separated by 10 Myr,
with no other massive star formation occurring during this interval.
 
 Tang \& Dauphas (2012) point out another possible explanation for
low $^{60}$Fe/$^{56}$Fe ratios and high $^{26}$Al/$^{27}$Al ratios:
$^{26}$Al is synthesized in the outer layers of massive stars, whereas
$^{60}$Fe is formed in the inner layers. If only the outer layers of
the progenitor star are ejected during the supernova explosion, or
if only the outer layers of the supernova shock are injected into
the presolar cloud, then the proper mixture of SLRIs might result
from a nearby supernova (e.g., Gritschneder et al. 2012). Boss \& Foster
(1998), however, found that the tail-end gas in a shock front
can be injected into a cloud core as efficiently as the leading-edge 
gas, casting some doubt on this explanation.

 Recent measurements of $^{41}$Ca/$^{40}$Ca ratios in CAIs (Liu et al. 
2012) have lowered the previously determined meteoritical ratio of 
$(1.41 \pm 0.14) \times 10^{-8}$ to a value of $\sim 4.2 \times 10^{-9}$,
about 10 times less than expected ISM ratios, which lie in the range
of $\sim 1.6 \times 10^{-8}$ to $\sim 6.4 \times 10^{-8}$ based on 
steady state galactic nucleosynthesis models. This finding implies 
that the Solar System's $^{41}$Ca may have decayed in, e.g., a GMC, 
for about 0.2 to 0.4 Myr prior to CAI formation (Liu et al. 2012), much 
shorter than the inferred free decay interval for $^{60}$Fe noted above.
Alternatively, the $^{41}$Ca may have been injected into the presolar 
cloud or solar nebula shortly before CAI formation (Liu et al. 2012).

 Given the latest results on the Solar System's $^{60}$Fe/$^{56}$Fe and
$^{41}$Ca/$^{40}$Ca ratios, it behooves us to reinvestigate the injection 
efficiencies of a wide variety of possible shock waves, ranging from 
massive star winds, to planetary nebulae, to classic supernovae shock 
waves, and to also reconsider how injection efficiencies might depend
on the location of SLRIs in the incoming shock wave. In this paper
we explore the effects of a wider range of shock wave and cloud core
properties than we have previously considered, including shock speed,
thickness, density, and temperature, and cloud core mass and rotation.
We restrict our exploration to 2D models, in order to more rapidly
explore the large parameter space of interest.

\section{Numerical Methods and Initial Conditions}

 The numerical calculations were performed with the FLASH2.5
adaptive mesh refinement (AMR) code (Fryxell et al. 2000).
The numerical methods and initial conditions are very much
the same as those used and described by the previous papers
in this series (Boss et al. 2008, 2010; Boss \& Keiser 2010, 2012),
and so will not be repeated here. Only a brief synopsis of the key 
details from Boss et al. (2008, 2010) will be given here, for 
completeness.
 
 As in Boss et al. (2008, 2010), we used the axisymmetric, 2D 
cylindrical coordinate ($R$, $Z$) version of FLASH2.5. For the 
standard models, the grid was 0.2 pc long in $Z$ (the direction 
of shock wave propagation) and 0.065 pc wide in $R$. The initial 
number of grid blocks was 5 in $R$ and 15 in $Z$, with each 2D block 
consisting of $8^2 = 64$ grid points; the maximum number of 
levels of grid refinement permitted was 5. Outflow boundary conditions
were used at all surfaces of the cylinder. Self-gravity was calculated
using the multipole expansion method, using Legendre polynomials
up to and including $l = 10$. 

 The standard target clouds were Bonnor-Ebert spheres of molecular 
hydrogen gas with initial central densities of $1.24 \times 10^{-18}$ 
g cm$^{-3}$, radii of 0.058 pc, and masses of 2.2 $M_\odot$, centered 
at $R = 0$ and $Z = 0.13$ pc. Bonnor-Ebert-like density profiles have 
been inferred for dense molecular cloud cores, such as Barnard 68
(Burkert \& Alves 2009) and L1696A (Ruoskanen et al. 2011). In the
Perseus molecular cloud, starless cores appear to have radial 
density profiles that are shallower than that of the singular 
isothermal sphere, where $\rho \propto r^{-2}$ (Schnee et al. 2010), and 
hence are roughly consistent with a Bonnor-Ebert-like initial profile.

 The standard initial shock 
had a shock gas density $\rho_s = 3.6 \times 10^{-20}$ g cm$^{-3}$ 
and shock width $w_s = 9.3 \times 10^{15}$ cm = 0.0030 pc. The mass
of the shock that was incident upon the target cloud was 0.017 $M_\odot$.
The post-shock wind had a density of $3.6 \times 10^{-22}$ g cm$^{-3}$.
Both the standard shock gas and the post-shock wind gas had initial
temperatures of 1000 K, while the target cloud and its ambient medium
started at 10 K, the latter with the same density as the post-shock wind.
Compressional heating and radiative cooling were included, using the
results of Neufeld \& Kaufman (1993) for cooling caused by optically 
thin molecular gas composed of H$_2$O, CO, and H$_2$. A color
field was defined to be non-zero initially in the shock front, in
order to be able to trace the degree to which shock front gas and
dust were injected into the collapsing protostellar cloud. In several
models, separate color fields were also initialized for the post-shock
wind and the pre-shock, ambient medium, to learn the fate of this 
material as well.

\section{Results}

 We present the results for a large range of variations in the
assumed parameters for the shock front and the target cloud. In each
case, we wish to compare the results for shock wave injection efficiencies
to those obtained with a standard model. Hence we begin with a summary 
of the results for the three standard models.

\subsection{Standard Models}

 Table 1 summarizes the three standard models (v10, v40, and v70), which 
are identical to the standard cloud and shock described above, differing
only in the assumed shock speeds: 10, 40, and 70 km s$^{-1}$, respectively. 
These models are similar to those presented by Boss et al. (2010), except
that the present models are for target clouds with twice the mass of
the Boss et al. (2010) clouds. Boss et al. (2010) found that shock
speeds in the range of 5--70 km s$^{-1}$ were able to simultaneously
trigger collapse and inject significant shock wave material: slower
shocks were unable to trigger collapse, while faster shocks shredded
the target clouds. Hence we focus here on the same range of shock speeds.

 Figure 1 depicts the evolution of model v40, which is typical in all
respects of the models where simultaneous triggered collapse and injection
occurs. The initial shock wave, moving at 40 km s$^{-1}$, strikes and 
compresses the edge of the target cloud, generating Rayleigh-Taylor fingers, 
leading to the onset of dynamic collapse along the symmetry axis
($R = x = 0$) and the formation of a protostar within $\sim 10^5$ yr.
Figure 1 shows that the protostar is accelerated by the shock front to
a speed of $\sim 1$ km s$^{-1}$. While significant, such speeds are
small with respect to the typical velocity dispersions seen for T Tauri
stars in Taurus and Orion, where dispersions of $\sim 5$ to 10 
km s$^{-1}$ are observed (Walter et al. 2000).

 Figure 2 shows the extent to which gas and dust from the shock wave, the
post-shock wind, and the ambient medium, are injected into the collapsing
protostar seen in Figure 1d. The velocity vectors show that while most
of this gas is blown downwind by the shock and post-shock wind, some
fraction of this material will be accreted by the protostar.

 Boss \& Keiser (2012) discussed several of the reasons why estimates of the 
injection efficiency are difficult to determine. Here we follow Boss \& Keiser 
(2012), and define the injection efficiency $f_i$ to be the fraction of the 
incident shock wave material that is injected into the collapsing cloud core. 
Table 1 then summarizes the amount of mass derived from the shock wave, post-shock
wind, and the ambient medium that is injected into the densest region of the 
collapsing protostar, i.e., regions with density greater than $10^{-16}$ g cm$^{-3}$, 
as used by Boss \& Keiser (2012). Table 1 shows that the injection
efficiencies for shock wave material are of order $f_i \sim 10^{-2}$ for
the standard models, as found by Boss \& Keiser (2010, 2012). In addition,
the amount of injected gas derived from the post-shock wind is typically
10 times higher, while that derived from the ambient medium (i.e., the pre-shock
gas) is typically over 10 times lower.

\subsection{Varied Target Cloud Densities and Radii}

 We first consider variations on standard model v40, where the density
and radius of the target cloud are changed. Figure 3 shows the result for a cloud 
that was shredded by a 40 km s$^{-1}$ shock and failed to collapse -- model 
v40-m1.9-r2, with a target cloud density 10 times lower than for model v40 (initial 
central density of $1.24 \times 10^{-19}$ g cm$^{-3}$) and a cloud radius twice 
as large (0.116 pc), leading to an initial cloud mass of 1.9 $M_\odot$.
Given that the shock in this model was identical 
to that in model v40 (Figures 1 and 2), it is clear that certain shock waves will 
only trigger the collapse of suitably dense initial cloud cores. In fact,
the same result -- a shredded cloud without sustained collapse -- was obtained
when the initial central density was decreased by only a factor of 5 to
$2.48 \times 10^{-19}$ g cm$^{-3}$ along with the doubled cloud radius,
for a cloud mass of 3.7 $M_\odot$. Shredding continued for these clouds
when the shock width was reduced by a factor of 10 to $w_s = 0.0003$ pc and 
the standard shock gas density ($\rho_s = 3.6 \times 10^{-20}$ g cm$^{-3}$) was
increased by factors of 10, 100, 200, 300, 400, and 1000. All of these models
refused to collapse, even when the computational volume was expanded
to 20 blocks in $Z = y$, the direction of shock wave propagation, to allow more 
time for collapse to begin before the cloud disappears off the lower edge
of the AMR grid.

 However, when the initial central density was decreased by a factor of 3.33 
to $3.72 \times 10^{-19}$ g cm$^{-3}$, along with the doubled cloud radius, 
raising the initial cloud mass to 5.4 $M_\odot$, dynamic collapse was 
achieved for the standard 40 km s$^{-1}$ shock (model v40-m5.4-r2). Table 2 
summarizes the injection efficiencies achieved for several of these models with  
varied shock densities and widths, all with the number of blocks in the $R = x$
direction doubled to 10. It can be seen that the values of $f_i$ are 
typically $\sim 10^{-4}$ to $\sim 10^{-3}$ for these 5.4 $M_\odot$ cloud models; 
however, $f_i = 0$ for model v40-m5.4-r2-w0.1-s400, as this model was compressed
to a maximum density of $\sim 10^{-16}$ g cm$^{-3}$, but was not quite driven 
into sustained collapse.
 
\subsection{Varied Target Cloud Rotation Rates}

 Table 3 summarizes the models with variations in the initial angular velocity
($\Omega_i$, in rad s$^{-1}$) about the $Z = y$ axis. All models had the standard
shock with $v_s$ = 40 km s$^{-1}$ and the standard 2.2 $M_\odot$ target cloud,
differing only in the initial rate of solid body rotation. The initial rotation
rates of $10^{-16}$ rad s$^{-1}$ to $10^{-12}$ rad s$^{-1}$ span the range
from essentially no rotation to the most rapid rotation rate possible for
collapse to occur for an isolated dense cloud core. Typical observed rotation
rates for dense cloud cores span the range of $\sim 10^{-14}$ rad s$^{-1}$ 
to $\sim 10^{-13}$ rad s$^{-1}$ (Goodman et al. 1993). 

 The rotation axis is the same as the symmetry axis, as that is the only possible 
configuration allowed for 2D, axisymmetric clouds. The effect of rotation will 
then be to flatten a collapsing cloud into the direction perpendicular to the 
rotation axis (i.e., outward along the $x = R$ axis), forming a rotationally 
flattened, protostellar disk that will be smacked face-on by the shock front. 
Such a configuration should lead to enhanced injection efficiencies by the
target cloud. Table 3 shows that, compared to the standard, non-rotating model 
v40, when cloud rotation is included, significantly higher injection efficiencies 
are obtained, as expected, by as much as a factor of $\sim$ 10. Figure 4 shows 
the results for four models with increasingly higher values of $\Omega_i$: as 
$\Omega_i$ increases, the collapsing protostar is accompanied by a progressively 
larger region with relatively high gas density (yellow-green regions), compared 
to the non-rotating model v40 (Figure 2a), allowing more shock wave material to be
intercepted in cloud gas that is likely to be accreted. Model v40-o13, in fact,
can be seen to have formed a structure similar to a large scale ($\sim$ 1000 AU) 
protostellar disk (Figure 4c). However, when $\Omega_i$ is increased to 
$10^{-12}$ rad s$^{-1}$, as in model v40-o12, the cloud does not collapse, 
and instead is shredded by the shock wave. Hence the extent to which initial 
cloud rotation can enhance the injection efficiency is limited in the end by 
the desire to also achieve high enough densities for dynamic collapse to ensue.
In addition, shock fronts which strike clouds where the rotation axis is not
aligned with the direction of shock propagation will not be able to
inject quite as much shock front matter, lowering the injection efficiencies.
The present results for aligned rotation axes and shock propagation directions
thus represent upper bounds for the possible enhancements due to rotation.
A fully 3D study will be needed to learn about injection efficiencies in
the case of non-aligned rotation axes.

\subsection{Varied Target Cloud Density Distribution}

 Table 4 summarizes the models with variations in the initial target cloud
density distribution, namely a perturbation to the Bonnor-Ebert sphere
with a binary perturbation along the symmetry ($Z = y$) axis, as
seen in Figure 5a. These models have a reference density of
$4.1 \times 10^{-19}$ g cm$^{-3}$, three times lower than the standard
target clouds, yielding an initial cloud mass of 1.0 $M_\odot$.
These clouds are so low in mass, in fact, that when evolved without
any shock wave, they expand and do not collapse. Hence these clouds
represent clumpy regions of the ISM that would not collapse on their
own, in contrast to the other, more massive target clouds, which 
eventually collapse on their own, even without being struck by a shock wave.

 Figure 5 shows that in spite of this basic difference, when struck by a
suitable shock wave, the 1.0 $M_\odot$ clouds can still be driven
into sustained gravitational collapse, in much the same way as the
other models. However, when $v_s$ is increased to 20 km s$^{-1}$ or
above, these low density clouds are shredded and do not collapse.
Table 4 summarizes these results, and shows that for the shock
speeds that do lead to collapse, the injection efficiencies can
be significant, i.e., $f_i$ as high as $5.2 \times 10^{-2}$ for 
model v10-m1.0.

\subsection{Varied Shock and Post-Shock Wind Temperatures}

 Table 5 summarizes the models with variations in the maximum shock ($T_s$)
and post-shock wind ($T_{ps}$) temperatures, compared to a standard shock
model with $T_s = T_{ps} = 1000$ K. All models collapse, except for
model v40-T100-s100, where the shock density was too high to result
in collapse, given the other parameters. The injection efficiencies
varied from $2.7 \times 10^{-4}$ to $1.8 \times 10^{-2}$ for the
$T_s = T_{ps} = 100$ K models, differing only in the value of $\rho_s$.
These models demonstrate that injection efficiencies continue to
be significant, even for considerably cooler shock and post-shock
temperatures than the standard values.

\subsection{Varied Shock Speeds, Densities, and Widths}

 Tables 6, 7, and 8 present the results for variations in the shock
densities ($\rho_s$) and widths ($w_s$), for shock speeds ($v_s$) of
10, 40, and 70 km s$^{-1}$, respectively. Many of these models
had extended grids ($N_y = 30$ or 40) in the shock direction ($Z = y$)
in order to allow the cloud to be followed downstream far enough
to determine if collapse would ensue. The models with $M_s = 0$ were
models where collapse did not occur. 
 
 Table 6 lists the models with $v_s = 20$ km s$^{-1}$, showing
that with a wide variety of shock densities and widths, injection
efficiencies of $\sim 10^{-2}$ to $\sim 10^{-1}$ can result if
collapse occurs. All of the 40 km s$^{-1}$ models listed in Table 7
were also described in Boss \& Keiser (2010), but with the
difference here that an improved methodology for estimating the
injection efficiencies (Boss \& Keiser 2012) is used here, which
usually results in higher estimates of $f_i$. Both estimates are
listed in Table 7 for ease of comparison. Finally, Table 8 shows
that for 70 km s$^{-1}$ shocks, injection efficiencies are still
significant, with $f_i \sim 10^{-3}$ to $\sim 10^{-2}$, but are
typically lower; e.g., $f_i$ = $7.4 \times 10^{-2}$, 
$5.0 \times 10^{-2}$, and $8.6 \times 10^{-3}$ for models 
v10-w0.1, v40-w0.1, and v70-w0.1, respectively.

\subsection{Discussion}

 We now turn to a discussion of which models might be the most appropriate
representations of observed stellar outflows and shock waves that might be 
carriers of freshly-synthesized SLRIs. Supernovae resulting from the 
core collapse of massive stars in the range of $\sim 20 - 60 M_\odot$ 
and planetary nebulae derived from intermediate-mass ($\sim 5 M_\odot$) 
AGB (Huss et al. 2009) or higher mass ($\sim 7 - 11 M_\odot$) super-AGB 
stars (Lugaro et al. 2012) have been proposed as possible sources 
of many Solar System SLRIs. In addition, significant amounts of 
$^{26}$Al are contained in the outflows of massive Wolf-Rayet stars, 
which are the predecessors to many Type II SN (Tatischeff, Duprat, 
\& de S\'er\'eville 2010). $^{26}$Al is abundant in the ISM:
Diehl et al. (2010) detected $\gamma$-rays from the Sco-Cen association
that implied the presence of $\sim 1.1 \times 10^{-4} M_\odot$ of
live $^{26}$Al. If this amount of $^{26}$Al pollutes a 
$\sim 10^{6} M_\odot$ GMC containing $^{27}$Al
at Solar System abundances (Anders \& Grevesse 1989), then the
GMC would have a ratio $^{26}$Al/$^{27}$Al $\sim 3 \times 10^{-6}$,
well below the inferred initial Solar System canonical level of 
$^{26}$Al/$^{27}$Al $\sim 5 \times 10^{-5}$ (MacPherson et al. 2012).

 The Cygnus Loop is a $\sim 10^4$ yr-old core collapse (Type II) supernova 
remnant (SNR) with $v_s \approx$ 170 km/sec and $w_s < 10^{15}$ cm 
(Blair et al. 1999), a width consistent with our thin shock models, 
with $0.1 w_s = 9.3 \times 10^{14}$ cm. W44 is also a core collapse SNR 
with $v_s = 20 - 30$ km/sec and a width less than $10^{16}$ cm 
(Reach et al. 2005), comparable to our standard shock width
$w_s = 9.3 \times 10^{15}$ cm. Hence SNR appear to be most consistent 
with our models with relatively high speed (70 km s$^{-1}$) shocks and 
$0.1 w_s$ or with models with lower speed (10 - 40 km s$^{-1}$) shocks 
and 1 or 0.1 $w_s$, respectively.

 The W44 SNR is expanding into gas with a number density $n \sim 10^2$ 
cm$^{-3}$ (Reach et al. 2005). Boss \& Keiser (2012) presented results for a 
3D version of our 2D model v40-w0.1-s400-20 (Table 7), a model with $v_s =$ 
40 km s$^{-1}$, shock number density $n_s \approx 4 \times 10^6$ cm$^{-3}$, 
and a shock width of $\approx 10^{15}$ cm. Boss \& Keiser (2012) noted 
that the shock density $n_s$ for an isothermal shock propagating in an ambient 
gas of density $n_{am}$ is $n_s/n_{am} = (v_s/c_{am})^2$, where $c_{am}$ is 
the ambient gas sound speed. For model v40-w0.1-s400-20, with $c_{am} = 0.2$ 
km s$^{-1}$ and $n_{am} = 10^2$ cm$^{-3}$, as in the case of the W44 SNR, the shock 
density should be $n_s = 4 \times 10^6$ cm$^{-3}$, the same as the density 
in model v40-w0.1-s400-20. Hence shock densities about 400 times higher than
in the standard shock model appear to be similar to what is expected to
be the case in realistic SNRs (Boss \& Keiser 2012).

 During the transition from an AGB star to a post-AGB star, low mass stars 
generate outflows known as planetary nebulae (van Marle \& Keppens 2012), 
with speeds of order 20 km s$^{-1}$. The planetary nebula Abell 39 is estimated 
to have a thickness of $\sim 3 \times 10^{17}$ cm (Jacoby, Ferland, 
\& Korista 2001), while planetary nebula PFP-1 is estimated to have 
a thickness of $\sim 5 \times 10^{17}$ cm (Pierce et al. 2004). These 
thicknesses are considerably larger than the standard shock
width $w_s = 9.3 \times 10^{15}$ cm, but are only a few times thicker
than the models with $10 w_s = 9.3 \times 10^{16}$ cm. Hence planetary 
nebulae appear to be most consistent with our models with low speed 
($\sim$ 10 km s$^{-1}$) shocks and large ($10 w_s$) shock widths.

 3D models of the interaction of a WR fast wind (initially moving at 
$2 \times 10^3$ km s$^{-1}$) with a previously emitted red supergiant
wind (moving at 10 - 15 km s$^{-1}$) led to the formation of R-T fingers
and expansion velocities of the interacting region less than 
100 km s$^{-1}$ (van Marle \& Keppen 2012). The thickness of the
shocked region was typically $\sim 10^{17}$ cm and the density
was $\sim 4 \times 10^{-22}$ g cm$^{-3}$. These quantities correspond to
shock widths and densities of $\sim$ 10 $w_s$ and 0.01 $\rho_s$,
respectively. The latter value is considerably lower than any of the values 
explored in the present study. Foster \& Boss (1996) found that
such a low-density shock would not have enough momentum to compress
a target cloud into collapse. Furthermore, the gas temperature was 
forced to remain above $10^4$ K everywhere in the van Marle \& Keppen (2012) 
simulations. WR winds are fully ionized and are driven by thermal
pressure. For these reasons a direct comparison to the present models is 
not possible. It is unclear what would happen when a WR wind shell
encounters a dense molecular cloud core, but the fact that the gas
is fully ionized and unable to cool below $10^4$ K implies that such a
hot shock would shred a target cloud to pieces (Foster \& Boss 1996). 

 The dilution factor $D$ is defined as the ratio of the amount of mass 
injected into the collapsing cloud that was derived from the shock 
wave ($M_s$) to the amount of mass derived from the target cloud. The Tables
list $M_s$ for each of the models. For a final system mass of 1 $M_\odot$,
the dilution factor is then simply $D = M_s/M_\odot$. However, because a
supernova shock starting at $\sim 2000$ km s$^{-1}$ must be slowed 
down by snowplowing a large amount of the ISM to reach a speed as
low as in these models, we follow Boss \& Keiser (2012) in
defining a third parameter, $\beta$, to be the ratio of shock front mass
originating in the supernova to the total mass in the shock, including that
swept up from the intervening ISM. In this case, the dilution factor 
becomes $D = \beta M_s$. For a SNR to slow down from 2000 km s$^{-1}$
to 10 - 70 km s$^{-1}$, as in our models, $\beta = 0.005 - 0.035$.
This means that the shock front must sweep up between $\sim 30$ and 
$\sim 200$ times its own mass before slowing to the proper speeds.
In the case of the W44 SNR, moving at $20 - 30$ km/sec, the shock
must have already swept up about 100 times its original mass. For a W44-like 
shock with $n_s = 4 \times 10^6$ cm$^{-3}$ and $w_s = 10^{15}$ cm, this
means that the amount of swept-up intervening gas was $\sim 99\%$
of the shock's current gas mass, i.e., a column depth (by number) of 
$\sim 4 \times 10^{21}$ cm$^{-2}$. For the gas that W44 is currently
expanding into, with a number density $\sim 10^2$ cm$^{-3}$ (Reach 
et al. 2005), this would require passing through $\sim$ 10 pc of
such intervening gas. The diameter of the W44 SNR is $\sim$ 11 pc
(Reach et al. 2005), in agreement with this estimate. Such number densities 
($\sim 10^2$ cm$^{-3}$) are typical of the average giant molecular cloud 
densities (Williams et al. 2000).

  The dilution factors needed to match inferred meteoritical abundances 
of SLRIs are uncertain, in part because the inferred initial abundances 
for some SLRIs are in flux (e.g., for $^{60}$Fe, as previously noted), 
but also because estimates of the production rates of SLRIs such as
$^{60}$Fe and $^{26}$Al are uncertain by large factors (Tur et al. 2010),
as noted by Boss \& Keiser (2012). Given these caveats, dilution factors 
for a supernova source range from $D \sim 10^{-4}$ to $10^{-3}$ 
(Takigawa et al. 2008) to $D = 3 \times 10^{-3}$ (Gaidos et al. 2009).
Given the $\beta$ factor, this means that a SN source requires models with:
1 $w_s$ and 10 km s$^{-1}$ ($\beta = 0.005$) to have $M_s = 0.02 - 0.6 M_\odot$; 
0.1 $w_s$ and 40 km s$^{-1}$ ($\beta = 0.02$) to have 
$M_s = 0.005 - 0.15 M_\odot$; or 0.1 $w_s$ and 70 km s$^{-1}$ 
($\beta = 0.035$) to have $M_s = 0.003 - 0.09 M_\odot$;  
in order to cover the range of $D \sim 10^{-4} - 3 \times 10^{-3}$. High
shock densities are also indicated for a SN source. The Tables show
that while none of the v10-series or v70-series models meet these requirements,
two models in Table 7 (v40-w0.1-s100 and v40-w0.1-s200)
reach $M_s = 0.0025$ and 0.0039 $M_\odot$, respectively, close
to the minimum amount needed. Given that rotation (Table 3) can 
increase the injection efficiencies by factors as large as 10, it
appears likely that a rotating target cloud may be able to meet
the required dilution factor for a $v_s = 40$ km s$^{-1}$ shock.

 For an AGB star, Trigo-Rodr\'iguez et al. (2009) estimate 
$D \sim 3 \times 10^{-3}$. Since $\beta \sim 1$ for an AGB source,
$D = M_s/M_\odot$, and an AGB source would require a model with 
$M_s \sim 3 \times 10^{-3} M_\odot$, 10 $w_s$, and $v_s \sim 20$ km s$^{-1}$.
The models with $v_s =$ 10 or 40 km s$^{-1}$ come the closest
to the correct shock speed. Table 6 shows that none of the
v10-w10-series succeeded in triggering collapse. Table 7 shows
that one of the v40-w10 series of models, v40-w10, was able to 
trigger collapse and inject $M_s \sim 4 \times 10^{-4} M_\odot$,
about a factor of 10 times too low to support the AGB star scenario.
However, the rotating target cloud models (Table 3) show that with
significant rotation, $M_s$ can be increased by a factor of 10, so
in that case an AGB star remains as a possibility.

 The Table 3 models show that target cloud rotation is likely to be a
key factor for achieving the dilution factors apparently
required for the Solar System's SLRIs. Tables 2 and 4 show that
target clouds with larger masses than the standard cloud are also
more likely to achieve collapse and significant injection. Table 5
shows that the maximum temperatures assumed for the shock and
post-shock gas can also have an effect on the outcome.
 
 Table 1 shows that more post-shock wind is likely to be accreted
by the collapsing protostar than shock wave matter, by as much
as a factor of 10. This is similar to the result found in the
2D models of Boss \& Foster (1998). Hence, these models suggest
that explaining the evidence for low $^{60}$Fe/$^{56}$Fe ratios and 
high $^{26}$Al/$^{27}$Al ratios as being due to their synthesis
in different layers of a massive star (Tang \& Dauphas 2012) 
will only work if the inner layers fall back onto the stellar
remnant, not if they are ejected and follow behind the leading
edge of the SN shock.

 Finally, we note that the present estimates strictly apply only to the
injection of shock wave gas carrying SLRIs, or dust grains small enough
to be carried along with the gas. Large grains can be injected with
higher efficiencies, as found by Ouellette et al. (2010). However,
such large grains appear to be rare in the ISM, and Boss \& Keiser
(2010) concluded that the overall effect should not be large. 
Nevertheless, to the extent that large grains carrying SLRIs are 
present in shock waves, the present estimates of SLRI injection
efficiencies should be considered to be lower bounds.
 
\section{Conclusions}

 We have shown that achieving simultaneous triggered collapse of a dense
molecular cloud core and injection of significant shock wave material
is by no means an easy task to achieve, at least not if the shock-cloud
interaction is required to inject sufficient SLRIs to match the
inferred abundances for the Solar System's most primitive meteoritical
components. Nevertheless, SN shock waves remain as the most likely 
stellar source of shock-injected SLRIs, though AGB stars remain
in the running, provided that the enhanced injection efficiencies
found for rotating target cloud cores can raise the dilution factors
to the required levels. AGB stars have a significant probability of
passing through a molecular cloud complex (Kastner \& Myers 1994),
and our models show that their winds can trigger cloud core collapse, 
so an AGB star source cannot be discounted on these grounds.
Wolf-Rayet star winds, on the other hand, do not appear to be 
suitable sources of the shock waves considered in this work, as their
hot, fully ionized winds are driven by thermal pressure, as opposed to
the momentum-driven, relatively cool winds shown here to be capable
of triggering collapse and injection. When collapse does occur,
significant post-shock wind gas and dust is likely to be
accreted by the growing protostar, adding to the isotopic brew obtained
from the stellar source of the shock wave.

 Future fully 3D models of this triggering and injection scenario
(e.g., Boss \& Keiser 2012) need to consider the case of rotating target clouds,
in order to boost the injection efficiencies and give this scenario the
best possible chance of remaining as a contender for explaining the origin of
the Solar System's SLRIs. Such models are currently in progress at DTM.

\acknowledgements

 We thank Roger Chevalier for discussions long ago, and the 
referee for suggesting several improvements to the manuscript.
The calculations were performed on the dc101 cluster at DTM.
This research was supported in part by NASA Origins of Solar Systems 
grant NNX09AF62G and is contributed in part to NASA Astrobiology 
Institute grant NNA09DA81A. The software used in this work was in 
part developed by the DOE-supported ASC/Alliances Center for 
Astrophysical Thermonuclear Flashes at the University of Chicago.

\clearpage
\begin{deluxetable}{lccccc}
\tablecaption{Results for the three standard 2.2 $M_\odot$ target cloud 
models with varied shock speeds ($v_s$ in km s$^{-1}$), showing the amount 
of mass in the collapsing protostar derived from the shock wave
($M_s$), post-shock wind ($M_{psw}$), and ambient medium
($M_{am}$), along with the injection efficiency ($f_i$) for the shock wave
material.
\label{tbl-1}}
\tablewidth{0pt}
\tablehead{\colhead{Model} 
& \colhead{$v_s$} 
& \colhead{$M_s/M_\odot$} 
& \colhead{$M_{psw}/M_\odot$} 
& \colhead{$M_{am}/M_\odot$} 
& \colhead{$f_i$} }
\startdata

v10 & 10 & $8.0 \times 10^{-5}$ & $1.5 \times 10^{-4}$ & $2.0 \times 10^{-6}$ & $4.7 \times 10^{-3}$ \\ 

v40 & 40 & $1.5 \times 10^{-4}$ & $1.1 \times 10^{-3}$ & $4.0 \times 10^{-6}$ & $8.8 \times 10^{-3}$ \\ 

v70 & 70 & $5.5 \times 10^{-4}$ & $4.6 \times 10^{-3}$ & $3.7 \times 10^{-5}$ & $3.2 \times 10^{-2}$ \\ 

\enddata
\end{deluxetable}

\clearpage
\begin{deluxetable}{lccccc}
\tablecaption{Results for the 5.4 $M_\odot$ target cloud models with varied
shock gas densities (in units of the standard shock density, $\rho_s = 3.6 
\times 10^{-20}$ g cm$^{-3}$) and shock widths (in units of the standard shock 
width, $w_s = 0.0030$ pc) at the standard shock speed of $v_s$ = 40 km s$^{-1}$.
$N_y$ is the number of blocks in the grid in the $y = Z$ direction. 
$M_s$ is the amount of mass in the collapsing protostar derived from the shock 
wave and $f_i$ is the injection efficiency for the shock wave material.
\label{tbl-2}}
\tablewidth{0pt}
\tablehead{\colhead{Model} 
& \colhead{$N_y$} 
& \colhead{$w_s$} 
& \colhead{$\rho_s$} 
& \colhead{$M_s/M_\odot$} 
& \colhead{$f_i$} }
\startdata

v40-m5.4-r2            & 20 & 1   & 1   & $4.5 \times 10^{-6}$ & $6.4 \times 10^{-5}$ \\ 

v40-m5.4-r2-w0.1-s250  & 30 & 0.1 & 250 & $3.0 \times 10^{-3}$ & $1.7 \times 10^{-3}$ \\ 

v40-m5.4-r2-w0.1-s300  & 30 & 0.1 & 300 & $1.3 \times 10^{-3}$ & $6.2 \times 10^{-4}$ \\ 

v40-m5.4-r2-w0.1-s400  & 30 & 0.1 & 400 &  0                   & 0                    \\ 

\enddata
\end{deluxetable}

\clearpage
\begin{deluxetable}{lccc}
\tablecaption{Results for the 2.2 $M_\odot$ target cloud models with varied initial 
rotation rates ($\Omega_i$, in rad s$^{-1}$) about the $Z = y$ axis, compared to 
the standard, non-rotating model v40. All models had $v_s$ = 40 km s$^{-1}$.
$M_s$ is the amount of mass in the collapsing protostar derived from the shock 
wave and $f_i$ is the injection efficiency for the shock wave material.
\label{tbl-3}}
\tablewidth{0pt}
\tablehead{\colhead{Model} 
& \colhead{$\Omega_i$} 
& \colhead{$M_s/M_\odot$} 
& \colhead{$f_i$} }
\startdata

v40         &  0.0          & $1.5 \times 10^{-4}$ & $8.8 \times 10^{-3}$ \\ 

v40-o16     & $10^{-16}$    & $4.7 \times 10^{-4}$ & $2.8 \times 10^{-3}$ \\ 

v40-o15     & $10^{-15}$    & $8.4 \times 10^{-4}$ & $4.9 \times 10^{-2}$ \\ 

v40-o14.5   & $10^{-14.5}$  & $1.0 \times 10^{-3}$ & $5.9 \times 10^{-2}$ \\ 

v40-o14     & $10^{-14}$    & $1.0 \times 10^{-3}$ & $5.9 \times 10^{-2}$ \\ 

v40-o13.5   & $10^{-13.5}$  & $1.3 \times 10^{-3}$ & $7.6 \times 10^{-2}$ \\ 

v40-o13     & $10^{-13}$    & $1.0 \times 10^{-3}$ & $5.9 \times 10^{-2}$ \\ 

v40-o12.5   & $10^{-12.5}$  & $7.8 \times 10^{-4}$ & $4.6 \times 10^{-2}$ \\ 

v40-o12     & $10^{-12}$    &  0               &  0               \\ 

\enddata
\end{deluxetable}

\clearpage
\begin{deluxetable}{lccc}
\tablecaption{Results for the 1.0 $M_\odot$ target cloud 
models with varied shock speeds ($v_s$ in km s$^{-1}$), showing the amount 
of mass in the collapsing protostar derived from the shock wave
($M_s$) and the injection efficiency ($f_i$) for the shock wave material.
\label{tbl-4}}
\tablewidth{0pt}
\tablehead{\colhead{Model} 
& \colhead{$v_s$} 
& \colhead{$M_s/M_\odot$} 
& \colhead{$f_i$} }
\startdata

v5-m1.0  &  5 & $9.3 \times 10^{-5}$ & $2.8 \times 10^{-3}$ \\ 

v10-m1.0 & 10 & $1.7 \times 10^{-3}$ & $5.2 \times 10^{-2}$ \\ 

v15-m1.0 & 15 & $4.6 \times 10^{-5}$ & $1.4 \times 10^{-3}$ \\ 

v20-m1.0 & 20 &  0                   & 0 \\ 

v40-m1.0 & 40 &  0                   & 0 \\ 

v70-m1.0 & 70 &  0                   & 0 \\ 

\enddata
\end{deluxetable}

\clearpage
\begin{deluxetable}{lccccc}
\tablecaption{Results for the 2.2 $M_\odot$ target cloud models with varied
shock ($T_s$ in K) and post-shock wind ($T_{ps}$ in K) temperatures and
varied shock gas densities (in units of the standard shock density, 
$\rho_s = 3.6 \times 10^{-20}$ g cm$^{-3}$), all for the standard shock 
speed of $v_s$ = 40 km s$^{-1}$. $M_s$ is the amount of mass in the 
collapsing protostar derived from the shock wave and $f_i$ is the 
injection efficiency.
\label{tbl-5}}
\tablewidth{0pt}
\tablehead{\colhead{Model} 
& \colhead{$T_s$} 
& \colhead{$T_{ps}$} 
& \colhead{$\rho_s$} 
& \colhead{$M_s/M_\odot$} 
& \colhead{$f_i$} }
\startdata

v40           & 1000 & 1000 & 1   & $1.5 \times 10^{-4}$ & $8.8 \times 10^{-3}$ \\ 

v40-T500      &  500 &  500 & 1   & $1.8 \times 10^{-4}$ & $9.2 \times 10^{-3}$ \\ 

v40-T100      &  100 &  100 & 1   & $5.2 \times 10^{-6}$ & $2.7 \times 10^{-4}$ \\ 

v40-T100-s10  &  100 &  100 & 10  & $3.5 \times 10^{-3}$ & $1.8 \times 10^{-2}$ \\ 

v40-T100-s100 &  100 &  100 & 100 &  0               &  0                       \\ 

\enddata
\end{deluxetable}

\clearpage
\begin{deluxetable}{lccccc}
\tablecaption{Results for the 2.2 $M_\odot$ target cloud models with varied
shock gas densities (in units of the standard shock density, $\rho_s = 3.6 
\times 10^{-20}$ g cm$^{-3}$) and shock widths (in units of the standard shock 
width, $w_s = 0.0030$ pc) at a shock speed of $v_s$ = 10 km s$^{-1}$. 
\label{tbl-6}}
\tablewidth{0pt}
\tablehead{\colhead{Model} 
& \colhead{$N_y$}
& \colhead{$w_s$} 
& \colhead{$\rho_s$} 
& \colhead{$M_s/M_\odot$} 
& \colhead{$f_i$} }
\startdata
v10         & 15 & 1   & 1   & $8.0 \times 10^{-5}$ & $4.7 \times 10^{-3}$ \\ 

v10-s0.1    & 15 & 1   & 0.1 & $1.1 \times 10^{-4}$ & $6.5 \times 10^{-2}$ \\ 

v10-s10-30  & 30 & 1   & 10  & $9.7 \times 10^{-4}$ & $5.7 \times 10^{-3}$ \\ 

v10-s100-40 & 40 & 1   & 100 & 0.0 & 0.0 \\

v10-w0.1          & 15 & 0.1 & 1    & $5.2 \times 10^{-5}$ & $7.4 \times 10^{-2}$ \\ 

v10-w0.1-s10      & 15 & 0.1 & 10   & $4.8 \times 10^{-4}$ & $6.9 \times 10^{-2}$ \\ 

v10-w0.1-s100-30  & 30 & 0.1 & 100  & $2.2 \times 10^{-3}$ & $3.1 \times 10^{-2}$ \\ 

v10-w0.1-s200-30  & 30 & 0.1 & 200  & $6.7 \times 10^{-3}$ & $4.8 \times 10^{-2}$ \\ 

v10-w0.1-s300-30  & 30 & 0.1 & 300  & $7.5 \times 10^{-3}$ & $3.4 \times 10^{-2}$ \\ 

v10-w0.1-s400-30  & 30 & 0.1 & 400  & $1.9 \times 10^{-2}$ & $6.8 \times 10^{-2}$ \\ 

v10-w0.1-s600-30  & 30 & 0.1 & 600  & $6.5 \times 10^{-3}$ & $1.5 \times 10^{-2}$ \\ 

v10-w0.1-s800-30  & 30 & 0.1 & 800  & $2.9 \times 10^{-2}$ & $4.9 \times 10^{-2}$ \\ 

v10-w0.1-s1000-30 & 30 & 0.1 & 1000 & $4.6 \times 10^{-2}$ & $6.3 \times 10^{-2}$ \\ 

v10-w0.1-s2000-30 & 30 & 0.1 & 2000 & $6.3 \times 10^{-3}$ & $4.3 \times 10^{-3}$ \\ 

v10-w0.1-s4000-30 & 30 & 0.1 & 4000 & $1.3 \times 10^{-2}$ & $4.5 \times 10^{-3}$ \\ 

v10-w10-s10-30   & 30 & 10 & 10   & 0.0 & 0.0 \\

v10-w10-s30-30   & 30 & 10 & 30   & 0.0 & 0.0 \\
 
v10-w10-s50-30   & 30 & 10 & 50   & 0.0 & 0.0 \\

v10-w10-s100-30  & 30 & 10 & 100  & $1.0 \times 10^{0}$ & $5.0 \times 10^{-2}$ \\ 

\enddata
\end{deluxetable}

\clearpage
\begin{deluxetable}{lcccccc}
\tablecaption{Results for the 2.2 $M_\odot$ target cloud models with varied
shock gas densities (in units of the standard shock density, $\rho_s = 3.6 
\times 10^{-20}$ g cm$^{-3}$) and shock widths (in units of the standard shock 
width, $w_s = 0.0030$ pc) at a shock speed of $v_s$ = 40 km s$^{-1}$. 
Injection efficiencies estimated with a different methodology by Boss \& Keiser
(2010) are listed in the last column.
\label{tbl-7}}
\tablewidth{0pt}
\tablehead{\colhead{Model} 
& \colhead{$N_y$}
& \colhead{$w_s$} 
& \colhead{$\rho_s$} 
& \colhead{$M_s/M_\odot$} 
& \colhead{$f_i$}
& \colhead{$f_i$ (BK2010)} }

\startdata

v40              & 15 & 1   & 1   & $1.5 \times 10^{-4}$  & $8.8 \times 10^{-3}$ & $1 \times 10^{-3}$ \\ 

v40-s0.1         & 15 & 1   & 0.1 & $5.3 \times 10^{-5}$ & $3.1 \times 10^{-2}$  & $6 \times 10^{-5}$ \\ 

v40-s10          & 15 & 1   & 10  & $6.4 \times 10^{-4}$ & $3.8 \times 10^{-3}$  & $3 \times 10^{-3}$ \\ 

v40-s100-30      & 30 & 1   & 100 & 0.0 & 0.0 & 0.0 \\
 
v40-w0.1         & 15 & 0.1 & 1   & $3.5 \times 10^{-5}$ & $5.0 \times 10^{-2}$ & $2 \times 10^{-4}$ \\ 

v40-w0.1-s10     & 15 & 0.1 & 10  & $4.8 \times 10^{-4}$ & $6.9 \times 10^{-2}$ & $2 \times 10^{-3}$ \\ 

v40-w0.1-s100    & 15 & 0.1 & 100 & $2.5 \times 10^{-3}$ & $3.6 \times 10^{-2}$ & $1 \times 10^{-2}$ \\ 

v40-w0.1-s200    & 15 & 0.1 & 200 & $3.9 \times 10^{-3}$ & $2.8 \times 10^{-2}$ & $2 \times 10^{-2}$ \\ 

v40-w0.1-s400-20 & 20 & 0.1 & 400 & $6.5 \times 10^{-4}$ & $2.3 \times 10^{-3}$ & $4 \times 10^{-2}$ \\ 

v40-w0.1-s800-20 & 20 & 0.1 & 800 & 0.0 & 0.0 & 0.0 \\

v40-w10          & 15 & 10  & 1   & $4.2 \times 10^{-4}$ & $2.2 \times 10^{-3}$ & $4 \times 10^{-4}$ \\ 

v40-w10-s10      & 15 & 10  & 10  & 0.0 & 0.0 & 0.0 \\

\enddata
\end{deluxetable}

\clearpage
\begin{deluxetable}{lccccc}
\tablecaption{Results for the 2.2 $M_\odot$ target cloud models with varied
shock gas densities (in units of the standard shock density, $\rho_s = 3.6 
\times 10^{-20}$ g cm$^{-3}$) and shock widths (in units of the standard shock 
width, $w_s = 0.0030$ pc) at a shock speed of $v_s$ = 70 km s$^{-1}$. 
\label{tbl-8}}
\tablewidth{0pt}
\tablehead{\colhead{Model} 
& \colhead{$N_y$}
& \colhead{$w_s$} 
& \colhead{$\rho_s$} 
& \colhead{$M_s/M_\odot$} 
& \colhead{$f_i$} }
\startdata

v70              & 15 & 1   & 1   & $5.5 \times 10^{-4}$ & $3.2 \times 10^{-2}$ \\ 

v70-s0.1         & 15 & 1   & 0.1  &$1.7 \times 10^{-5}$ & $1.0 \times 10^{-2}$ \\ 

v70-s10          & 15 & 1   & 10   & 0.0 & 0.0 \\

v70-s100-30      & 30 & 1   & 100  & 0.0 & 0.0 \\

v70-w0.1         & 15 & 0.1   & 1   & $6.0 \times 10^{-6}$ & $8.6 \times 10^{-3}$ \\ 

v70-w0.1-s10     & 15 & 0.1   & 10  & $6.8 \times 10^{-6}$ & $8.9 \times 10^{-4}$ \\ 

v70-w0.1-s100    & 15 & 0.1   & 100 & $3.9 \times 10^{-4}$ & $5.6 \times 10^{-3}$ \\ 

v70-w0.1-s200    & 15 & 0.1   & 200 & 0.0 & 0.0 \\

v70-w0.1-s200-20 & 20 & 0.1   & 200 & 0.0 & 0.0 \\

v70-w0.1-s400-20 & 20 & 0.1   & 400 & 0.0 & 0.0 \\

v70-w10          & 15 & 10   & 1    & $5.8 \times 10^{-4}$ & $3.1 \times 10^{-3}$ \\ 

v70-w10-s10      & 15 & 10   & 10   & 0.0 & 0.0 \\

\enddata
\end{deluxetable}

\clearpage

\begin{figure}
\vspace{-0.20in}
\plotone{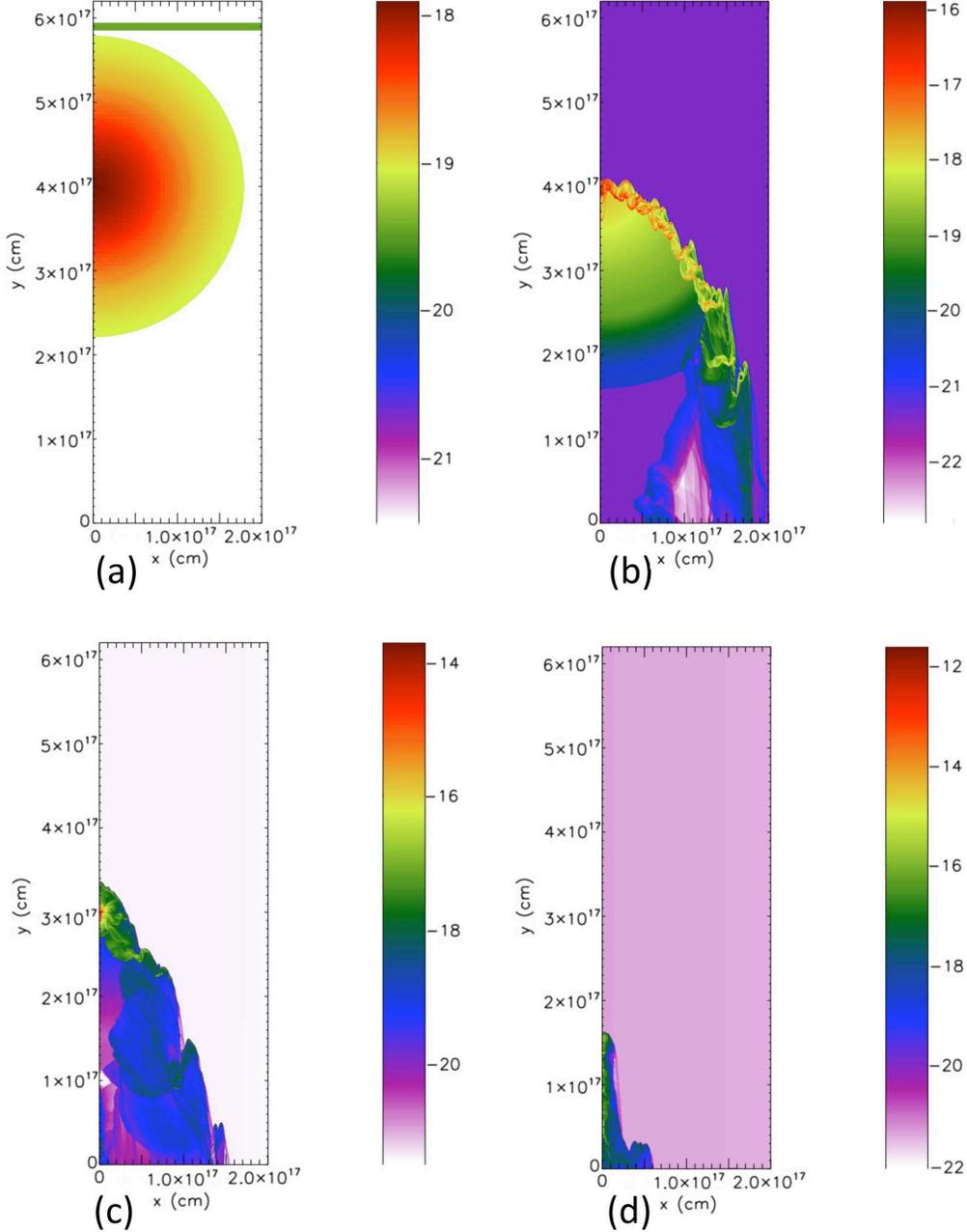}
\vspace{-1.5in}
\caption{Time evolution of model v40, showing the log of the
density distribution at four times: a) 0.0 yr, b) $3.34 \times 10^4$ yr,
c) $6.58 \times 10^4$ yr, and d) $10.4 \times 10^4$ yr. The initial (a) 
shock wave (green) moves downward at 40 km s$^{-1}$ to strike and compress
the target cloud, generating Rayleigh-Taylor fingers (b), leading to
the onset of dynamic collapse (c-d) and the formation of a protostar.
The 2D AMR code symmetry axis is along the left hand side of the plot. 
The $R = x$ axis is horizontal and the $Z = y$ axis is vertical.}
\end{figure}

\clearpage

\begin{figure}
\vspace{-0.5in}
\plotone{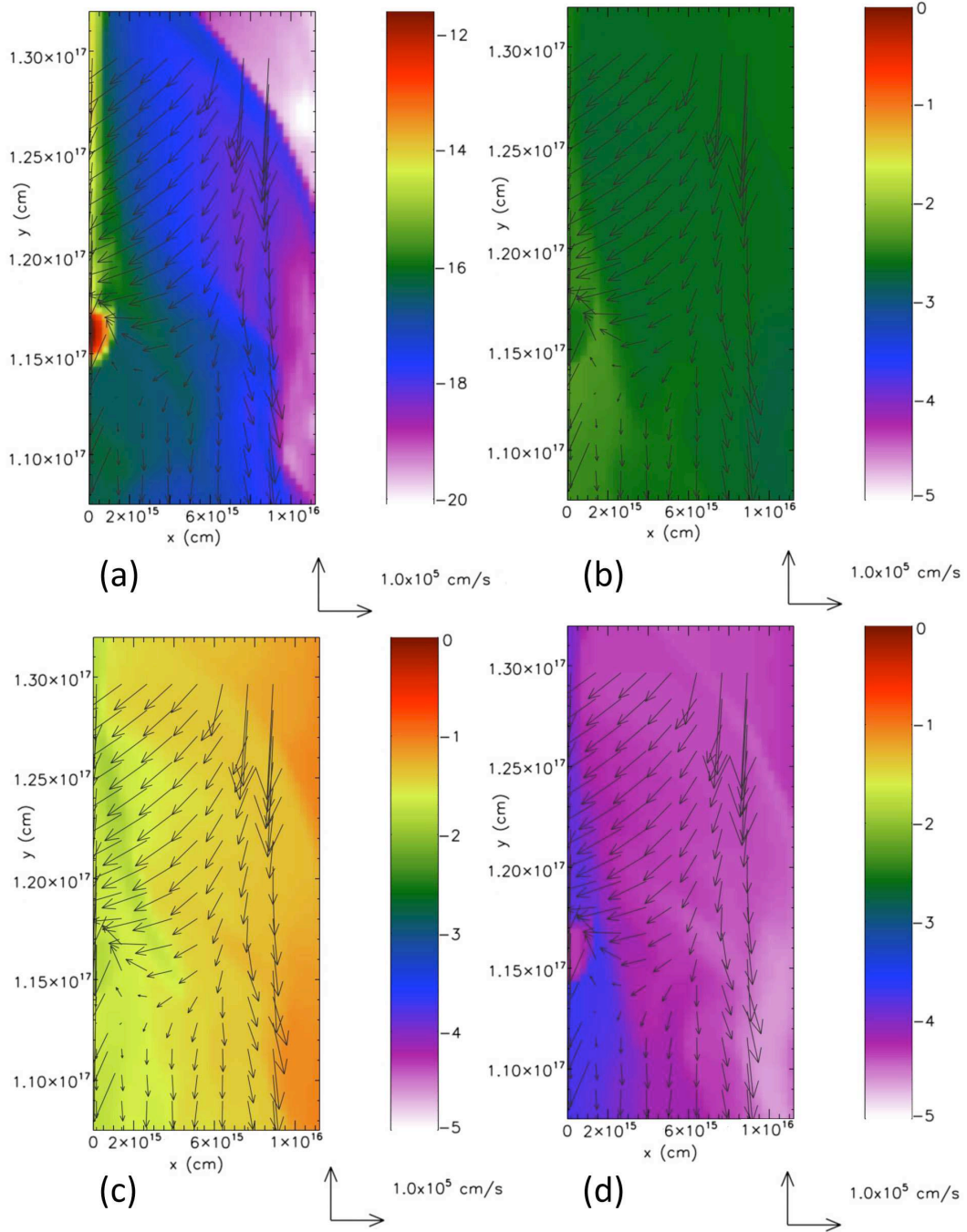}
\vspace{-1.5in}
\caption{Model v40 after $10.4 \times 10^4$ yr, plotted in the same manner as
in Figure 1, for four fields: a) log of the total gas density, b) log shock wave
gas density, c) log post-shock wind gas density, and d) log ambient medium gas
density. Velocity vectors are plotted for every fourth grid point in each 
direction, with the scale bar denoting 1 km s$^{-1}$.}
\end{figure}

\clearpage

\begin{figure}
\vspace{-0.20in}
\plotone{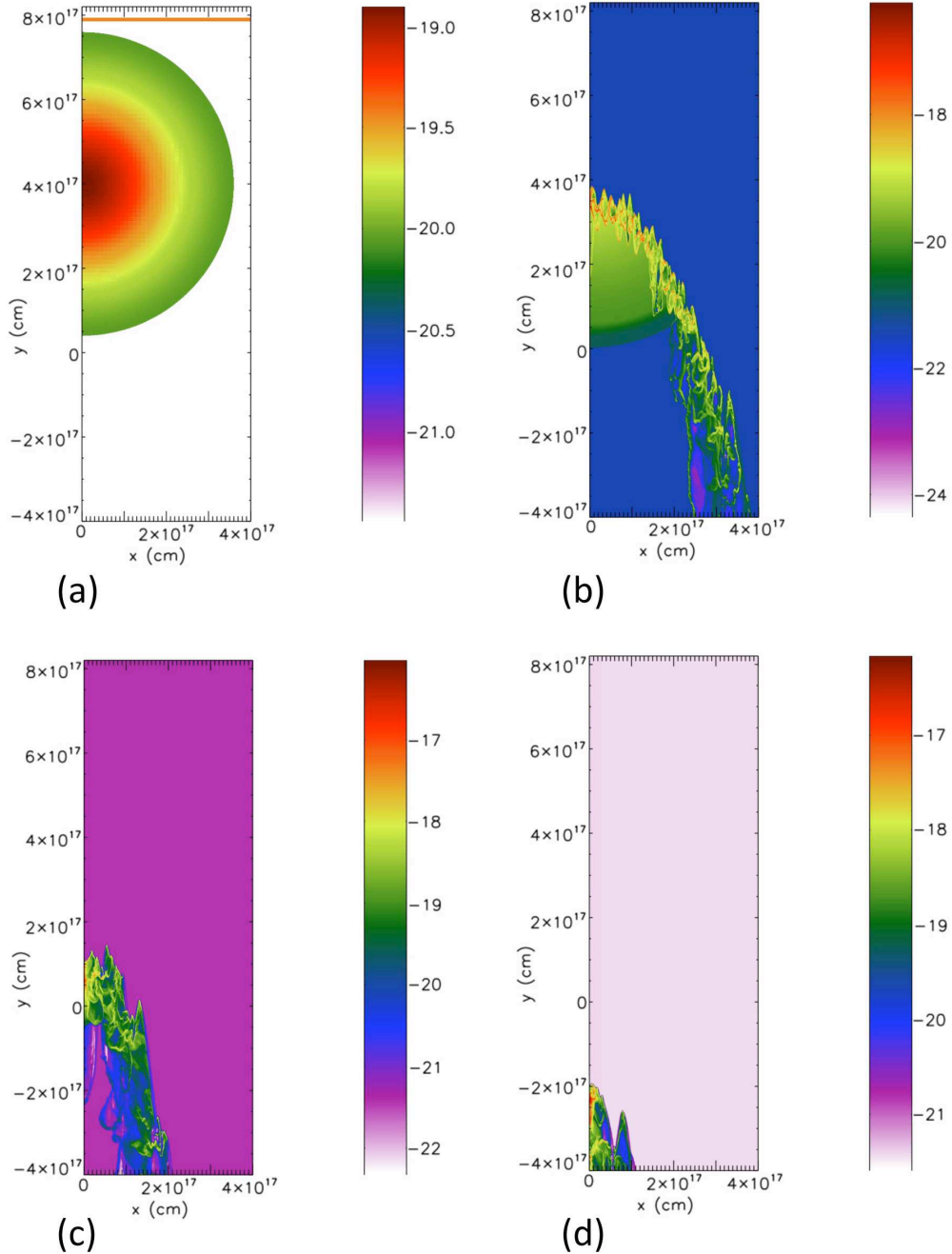}
\vspace{-1.5in}
\caption{Time evolution of model v40-m1.9-r2, plotted as in Figure 1, at times: 
a) 0.0 yr, b) $3.00 \times 10^4$ yr, c) $6.00 \times 10^4$ yr, and 
d) $8.99 \times 10^4$ yr. With a target cloud density 10 times lower
than for model v40 (Figures 1 and 2) and a cloud radius twice as large, 
the same standard 40 km s$^{-1}$ shock smacks the cloud b) and shreds 
it c), without inducing sustained gravitational collapse. The maximum 
density reached never exceeds $\sim 10^{-16}$ g cm$^{-3}$.}
\end{figure}

\clearpage

\begin{figure}
\vspace{-0.20in}
\plotone{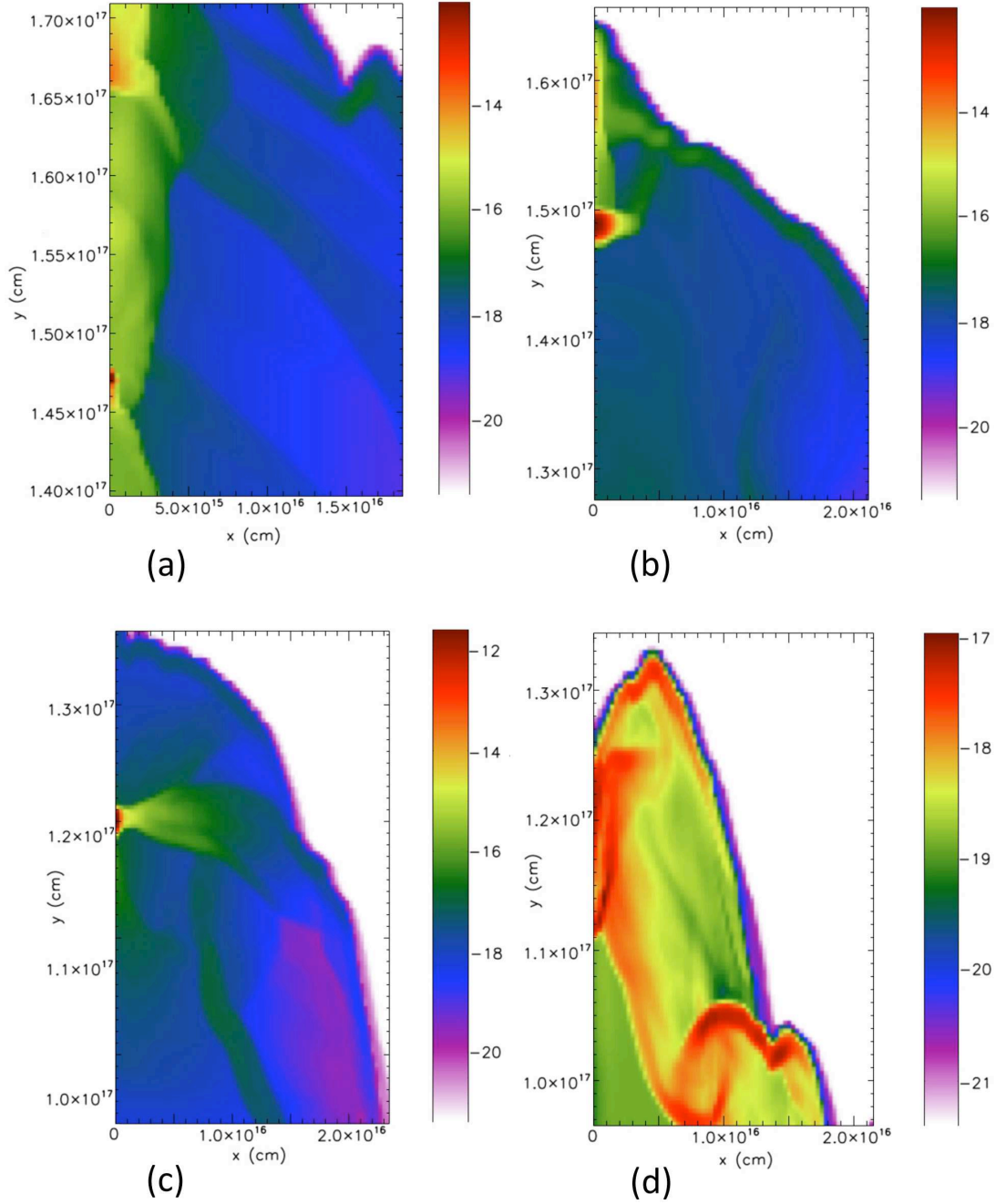}
\vspace{-1.5in}
\caption{Results of models with progressively larger initial cloud 
rotation rates about the $Z = y$ axis, showing the log of the gas density,
plotted as in Figure 1: a) model v40-o15 at $1.00 \times 10^5$ yr, 
b) model v40-o14 at $1.05 \times 10^5$ yr, c) model v40-o13 at 
$1.10 \times 10^5$ yr, and d) model v40-o12 at $8.99 \times 10^4$ yr. 
Model v40-o13 collapses and forms an extended protostellar disk, while 
model v40-o12 is so distended by the rotation that it fails to form a 
protostar, and is shredded instead.}
\end{figure}

\clearpage

\begin{figure}
\vspace{-0.20in}
\plotone{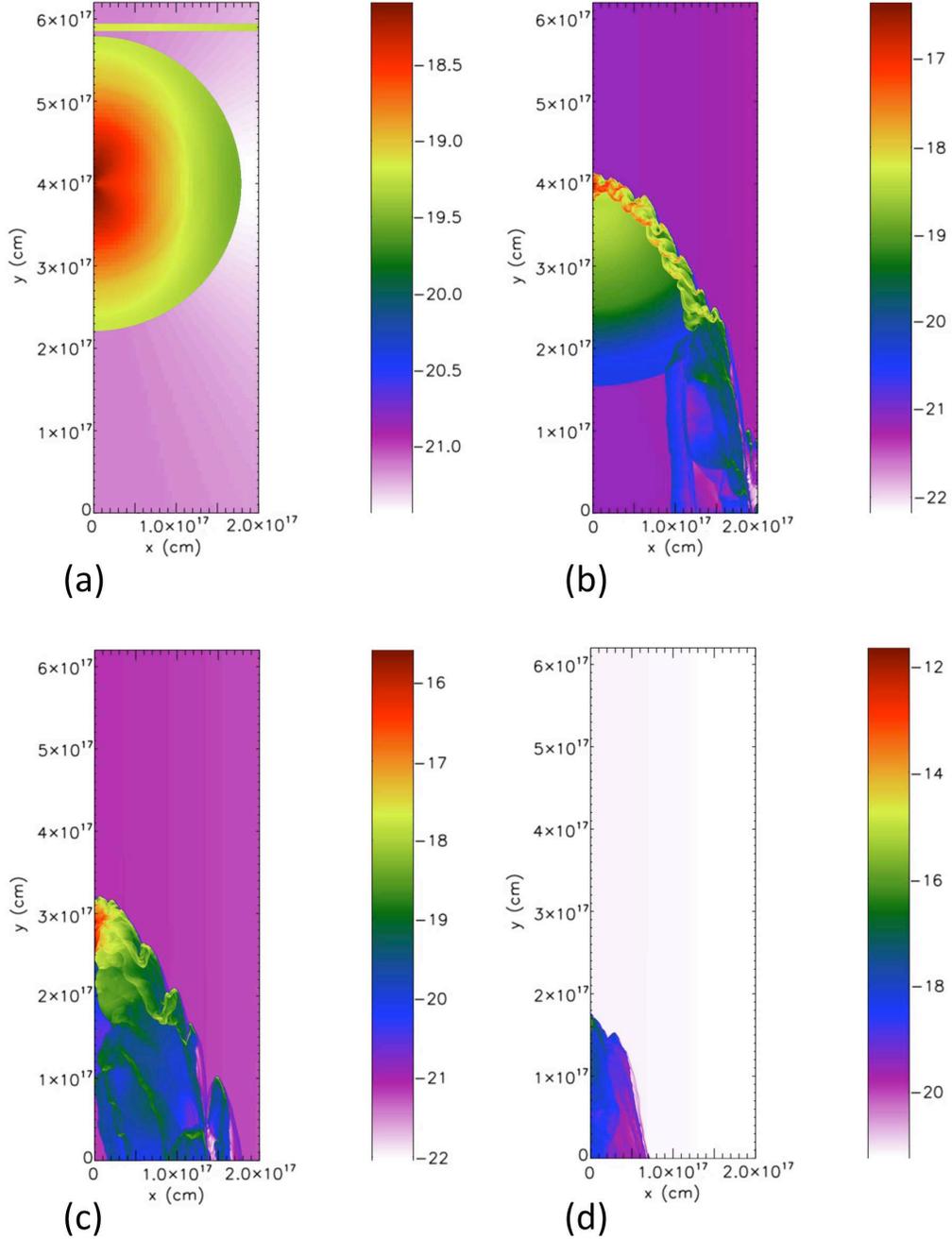}
\vspace{-1.5in}
\caption{Time evolution of model v15-m1.0, showing the log of the
density distribution at four times: a) 0.0 yr, b) $4.47 \times 10^4$ yr,
c) $9.02 \times 10^4$ yr, and d) $14.0 \times 10^4$ yr, plotted as
in Figure 1.}
\end{figure}

\end{document}